\documentclass[fleqn,usenatbib]{mnras}

\usepackage{newtxtext,newtxmath}

\usepackage[T1]{fontenc}

\DeclareRobustCommand{\VAN}[3]{#2}
\let\VANthebibliography\thebibliography
\def\thebibliography{\DeclareRobustCommand{\VAN}[3]{##3}\VANthebibliography}

\usepackage{graphicx}	
\usepackage{amsmath}

\title[Tilt Instability of Retrograde Discs]{Retrograde discs around one component of a binary are unstable to tilting}

\author[M. Overton et al.]{
Madeline Overton$^{1,2}$\thanks{E-mail: madeline.overton@unlv.edu}, Rebecca G. Martin$^{1,2}$, Stephen H. Lubow$^3$ and Stephen Lepp$^{1,2}$
\\
$^{1}$Nevada Center for Astrophysics, University of Nevada, Las Vegas, 4505 South Maryland Parkway, Las Vegas, NV 89154, USA\\
$^{2}$Department of Physics and Astronomy, University of Nevada, Las Vegas 4505 South Maryland Parkway, Las Vegas, NV 89154, USA\\
$^{3}$Space Telescope Science Institute, 3700 San Martin Drive, Baltimore, MD 21218, USA
}

\date{Accepted XXX. Received YYY; in original form ZZZ}

\pubyear{2023}

\begin{document}
\label{firstpage}
\pagerange{\pageref{firstpage}--\pageref{lastpage}}
\maketitle

\begin{abstract}
With hydrodynamic simulations we show that a coplanar disc around one component of a binary can be unstable to global tilting when the disc orbits in a retrograde direction relative to the binary. The disc experiences the largest inclination growth relative to the binary orbit in the outermost radii of the disc, closest to the companion. This tilt instability also occurs for test particles. A retrograde disc is much larger than a prograde disc since it is not tidally truncated and instead spreads outwards to the orbit of the companion. The coplanar retrograde disc remains circular while a coplanar prograde disc can become eccentric. We suggest that the inclination instability is due to a disc resonance caused by the interaction of the tilt with the tidal field of the binary. This model is applicable to  Be/X-ray binaries in which the Be star disc may be retrograde relative to the binary orbit if there was a sufficiently strong kick from the supernova that formed the neutron star companion. The accretion on to the neutron star and the resulting X-ray outbursts are weaker in the retrograde case compared to the prograde case.

\end{abstract}




\section{Introduction}

Be/X-ray binary systems consist of a high mass Be-type main sequence star with a compact object companion. The companion is often a neutron star \citep{Reig2011}, but may also be a black hole \citep{MunarAdrover2014}. The Be star rapidly rotates \citep{Slettebak1982,Porter1996,Porter2003} and possesses a viscous Keplarian decretion disc of material flowing out from the star \citep{Pringle1991,Lee1991,Hanuschik1996,Carciofi2011}.
Observations show that decretion discs that are tilted relative to the orbital plane of the binary are common \citep[e.g.][]{Hummel1998,Hughes1999,Hirata2007,martin2011be}. The tilt is likely due to a kick from the supernova from which the companion formed \citep{Brandt1995,Martinetal2009b,Salvesen2020}.

Material in the Be star disc may be captured by the companion neutron star and accreted onto its surface. Accretion results in two types of X-ray outbursts \citep{Stella1986,Negueruela1998}. Type~I outbursts occur each binary orbital period. For an eccentric orbit binary, this occurs when the companion approaches periastron \citep[e.g.][]{Negueruela2001,Negueruela2001b,Okazaki2002,Okazaki2013}, while for a circular binary, this can occur when the companion approaches the apastron of an eccentric disc \citep{Franchini2019b} at which point the companion is closest to the disc material. 
Type II outbursts are a few orders of magnitude brighter than type~I outbursts and they last for several orbital periods \citep{Kretschmar2013}. Type II outbursts can be the result of large eccentricity growth in the circumprimary disc \citep{Martinetal2014,Martin2019kl,Suffak2022}. This can occur through von Zeipel-Kozai-Lidov \citep[ZKL,][]{vonZeipel1910,Kozai1962,Lidov1962} oscillations in a highly misaligned disc around one component of a binary \citep{Martinetal2014b,Fu2015,Dogan2015,Lubow2017,Zanazzi2017,Franchini2019}.

Observations of the spin rate of neutron stars in Be/X-ray binaries have shown roughly equal numbers of spin up and spin down cases \citep{Klus2014} and it has been suggested that retrograde accretion may be common \citep{Christodoulou17}.
Retrograde circumprimary discs may also occur in merging supermassive binary black hole systems \citep[e.g.][]{King2006,Nixonetal2011a,Pereiraetal2019}. However, we are concerned with a time far before the merger phase where the companion drifts very slowly. There is also likely to be a circumbinary disc. Most simulations of mergers involving a retrograde disc have concentrated on the circumbinary disc \citep[e.g.][]{Nixon2015,Bankert2015}. Retrograde particle orbits around one component of a binary have been considered  previously and have been found to be stable closer to the companion than prograde orbits \citep{Morais2012} however, the orbits have been restricted to be in the binary orbital plane \citep[e.g.][]{Morais2022}.

Motivated by the observations of Be/X-ray binaries, in this Letter, for the first time we explore the dynamics of a retrograde  disc around one component of a binary. In section \ref{sec:methods} we describe and compare the numerical experiments of Be/X-ray binaries with prograde and retrograde circumprimary discs.  We find an inclination instability in retrograde discs that has not been seen before. To aid our understanding, in Section~\ref{sec:test_part}  we also investigate the behavior of test particles at key radii in the disc and discuss resonances in retrograde discs. We suggest that the inclination instability can be explained by a retrograde Lindblad resonance acting over the outermost regions of the disc, which arises from interaction of a small tilt of the disc away from the binary orbital plane with the tidal field of the binary. We draw our conclusions in Section~\ref{concs}.

\section{Hydrodynamical Simulations}
\label{sec:methods}

In this section we first discuss the setup of the hydrodynamical disc simulations and then the results of prograde and retrograde disc simulations. 

\subsection{Simulation setup}
\label{sec:SPH_sims} 

We use the {\sc phantom} \citep{Price2010,Lodato2010,Price2018}  smoothed particle hydrodynamics (SPH) code \citep{Monaghan1992,Bateetal1995,Price2012a} to model prograde and retrograde discs in Be/X-ray binary star systems. The binary system consists of a $M_\star= 18 \, \rm M_{\odot}$ Be star and a $M_{\rm NS}=1.4 \, \rm M_{\odot}$ neutron star companion, each  represented by a sink particle. Any particles that move inside the sink radius are accreted and their mass and angular momentum are added to the sink particle \citep{Bateetal1995}. The Be star radius is $r_* = 8 \, \rm R_{\odot}$ and the neutron star radius is $r_{\rm NS}= 1 \, \rm R_{\odot}$, therefore, we do not model the inner region of the neutron star’s accretion disc. The binary orbit is initially circular with a separation of $a=95\,\rm R_\odot$, which gives an orbital period of $P_{\rm orb}=24.3 \,$days. These parameters are motivated by Be/X-ray binaries with nearly circular orbits such as 2S 1553–542, which has an orbital eccentricity $e < 0.09$ and an orbital period of 30.6 days \citep{Reig2007b} as well as the Be/X-ray binary transient 4U 0115+63 which has an orbital period of 24.3 days \citep{Giacconi1972, Rappaport1978}. 

We initialize a circular disc of total mass $M_{\rm di}=10 ^{-8} \, \rm M_{\odot}$ around the Be star with $5 \times 10^5$ particles. The gas particles are initially in Keplerian orbits with a small correction due to pressure effects. The disc initially extends from $R=8$ to $30 \, \rm R_{\odot}$. The surface density of the disc initially follows a profile $\Sigma =\Sigma_0 (r_\star/R)^{3/2}[1-(r_\star/R)^{1/2}]$, where $\Sigma_0$ is a scaling constant.
In the vertical $z$ direction, the particles are initially distributed with a particle density $\rho \propto \exp(-z^2/(2H^2))$, where $H$ is the disc scale height.
The later disc evolution is not very sensitive to the initial surface density profile because the surface density evolves significantly during the simulation.  
The simulations are independent of the total disc mass and, by extension, $\Sigma_0$ since the disc mass is much smaller than the binary mass, and we do not include self-gravity of the disc.
For example, if the initial total disc mass was increased by some factor while keeping all other parameters fixed, then the surface density profiles would increase by the same factor at all times. 
This property holds because we apply a locally isothermal equation of state so that the disc temperature is independent of density. However, the disc density would have an effect on the predicted observables, including the accretion rates and the resulting X-ray flux from outbursts.

We adopt a locally isothermal equation of state of the form  
\begin{equation}
      \label{c_s}
      c_{\rm s} = c_{\rm s0} \bigg( \frac{M_*}{R} + \frac{M_{\rm NS}}{R_{\rm NS}} \bigg)^{q} 
\end{equation}
\citep{Farris2014} with $q=0.5$, where $R$ and $R_{\rm NS}$ are the distances from the primary and secondary objects, respectively. The disc temperature scales with $T\propto c_{\rm s}^2$. The constant of proportionality $c_{\rm s0}$ is chosen to give the disc aspect ratio, $H/R$, at the inner disc radius ($R = r_* = 8 \, \rm R_{\odot}$) to be $0.01$. This sound speed gives 
$c_{\rm s} \propto R^{-q}$ close to the Be star and  $c_{\rm s} \propto R_{\rm NS}^{-q}$ close to the neutron star.  The sound speed is $c_{\rm s} = H \Omega$, where the Keplerian frequency is $\Omega=\sqrt{GM_*/R^3}$ and therefore the circumprimary disc has an nearly constant disc aspect ratio with radius of $H/R \approx0.01$. These choices give an approximate  initial mean smoothing length over the disc scale height of $\left<h\right>/H  = 0.9$.

The \cite{SS1973} viscosity parameter $\alpha \approx 0.3$ in Be star discs \citep[e.g.][]{Rimulo2018, Ghoreyshi2018,Martin2019}. 
We adapt the SPH artificial viscosity according to the recommended procedure given by \cite{Lodato2010}. We use an artificial viscosity parameter of $\alpha^{\rm AV} = 3.32$. We can calculate an approximate \cite{SS1973} viscosity parameter with 
\begin{equation}
    \alpha \approx \frac{1}{10}\alpha_{\rm AV}\frac{\left<h\right>}{H}
\end{equation}
\citep[e.g.][]{Okazaki2002}.
In addition, we take $\beta^{\rm AV} = 2$, which correspond to the non-linear component of the artificial viscosity that prevents particle penetration in high Mach number flows \citep{Monaghan1989}.

We consider two initial inclinations for the circumprimary disc with respect to the binary orbital plane, $i=0^\circ$ (prograde) and $180^{\circ}$ (retrograde). This is the only difference between the setup of the prograde and retrograde cases.
We run the simulations for a time of $50 \,\rm P_{\rm orb}$. The binary and  circumprimary disc parameters are summarized in Table~\ref{tab:param_table}.

\begin{table}
	\centering
	\caption{Parameters for the binary and the circumprimary disc.}
	\label{tab:param_table}
	\begin{tabular}{lccr} 
		\hline
		  Binary parameters & Symbol/Units & Value \\
		\hline
		  Mass of Be Star & $M_*/ M_{\odot}$ & 18  \\
            Sink Radius of Be Star & $r_* / R_{\odot}$ & 8 \\
		  Mass of NS companion & $M_{\rm NS}/ M_{\odot}$ & 1.4 \\
            Sink radius of NS & $r_{\rm NS} / R_{\odot}$ & 1 \\
            Orbital period & $P_{\rm orb}/ \rm day$ & 24.3 \\
            Semi-major axis & $a/R_{\odot}$ & 95 \\
            Eccentricity & $e$ & 0 \\
            \hline 
            Initial Circumprimary Disc Parameters & Symbol & Value \\
            \hline 
		  Disc mass  & $M_{\rm di}/M_{\odot}$ & $10^{-8}$ \\
            Inclination to the binary orbital plane & $i$ & $0$, $180^{\circ}$ \\
            Initial disc outer radius & $R/R_{\odot}$ & 30 \\
            Shakura \& Sunyaev viscosity parameter & $\alpha$ & 0.25-0.39 \\
            Aspect ratio & $H/R \, (R = r_*)$ & 0.01 \\
		\hline
	\end{tabular}
\end{table}

There is no addition of mass to the disc during the simulation and therefore we model an accretion disc rather than a decretion disc. However, the outer parts of the disc behave like a decretion disc as they spread outwards \citep[e.g.][]{Pringle1981,MartinandLubow2011}. This difference does not play a role in the tilt instability that we discuss. Addition of material would occur close to the Be star in the very inner parts of the disc  \citep[e.g.][]{Okazaki2002,Suffak2022} that are not affected by the tilt instability that we later describe. The disc mass decreases during the simulation as particles can be accreted on to either star and ejected to form circumbinary material. 

\subsection{Prograde disc}

\begin{figure*}
    \centering
    \includegraphics[width=0.8\textwidth]{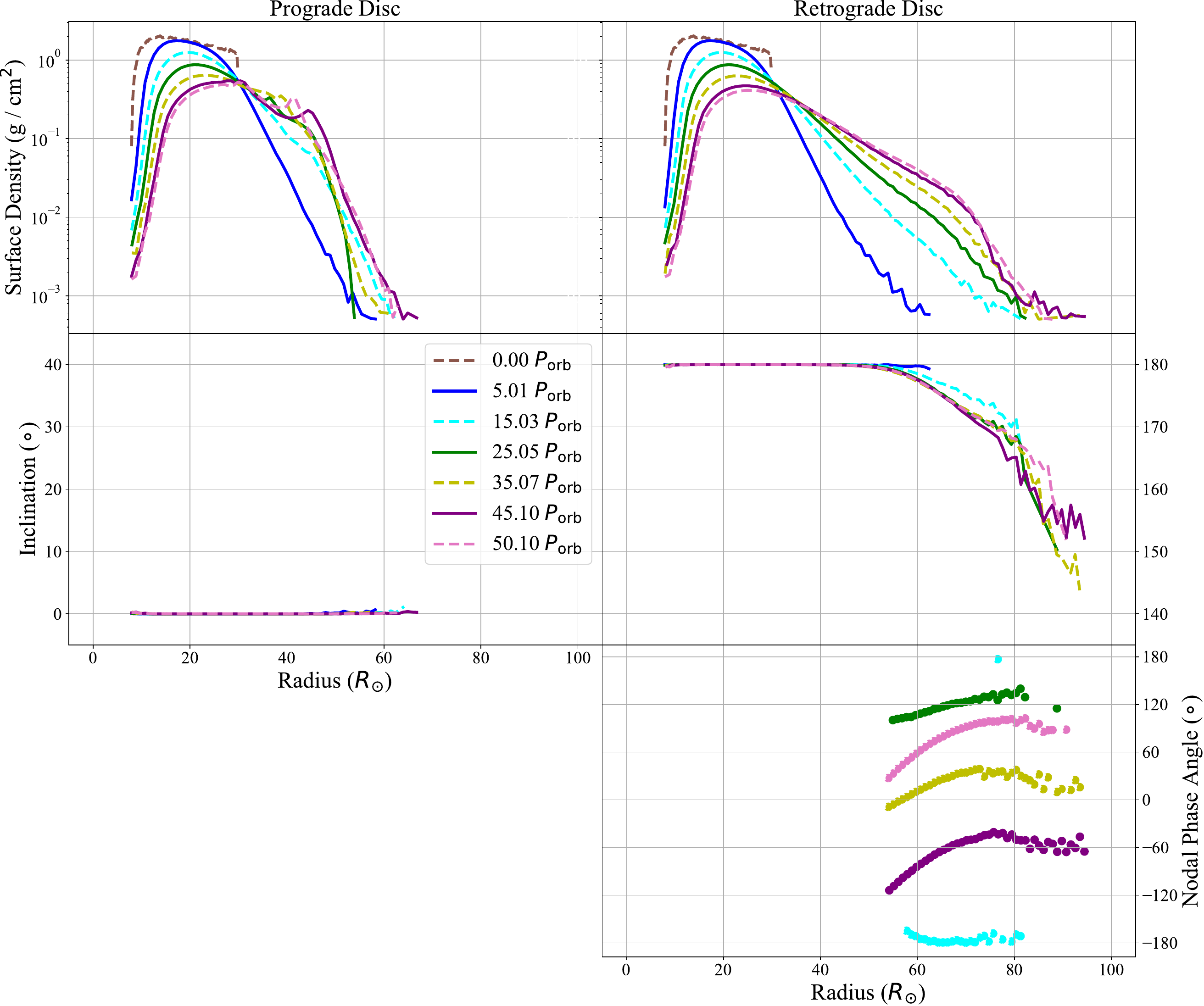} 
    \caption{Surface density (upper panels), inclination to the binary orbital plane (middle panels) and nodal phase angle (lower right panel) of the prograde (left panels) and retrograde (right panels) discs as a function of radius at different times in the simulations. 
    The surface densities are shown where they are greater than $5 \times 10^{-4} \, \rm g \ cm^{-2}$.
    The nodal phase angle is not well defined for a coplanar disc, so we only plot the values for the portions of the disc with inclination  $i<179^\circ$ for a retrograde disc. }
    \label{fig:sigma_i}
\end{figure*}

\begin{figure*}
    \includegraphics[width=0.4\textwidth]{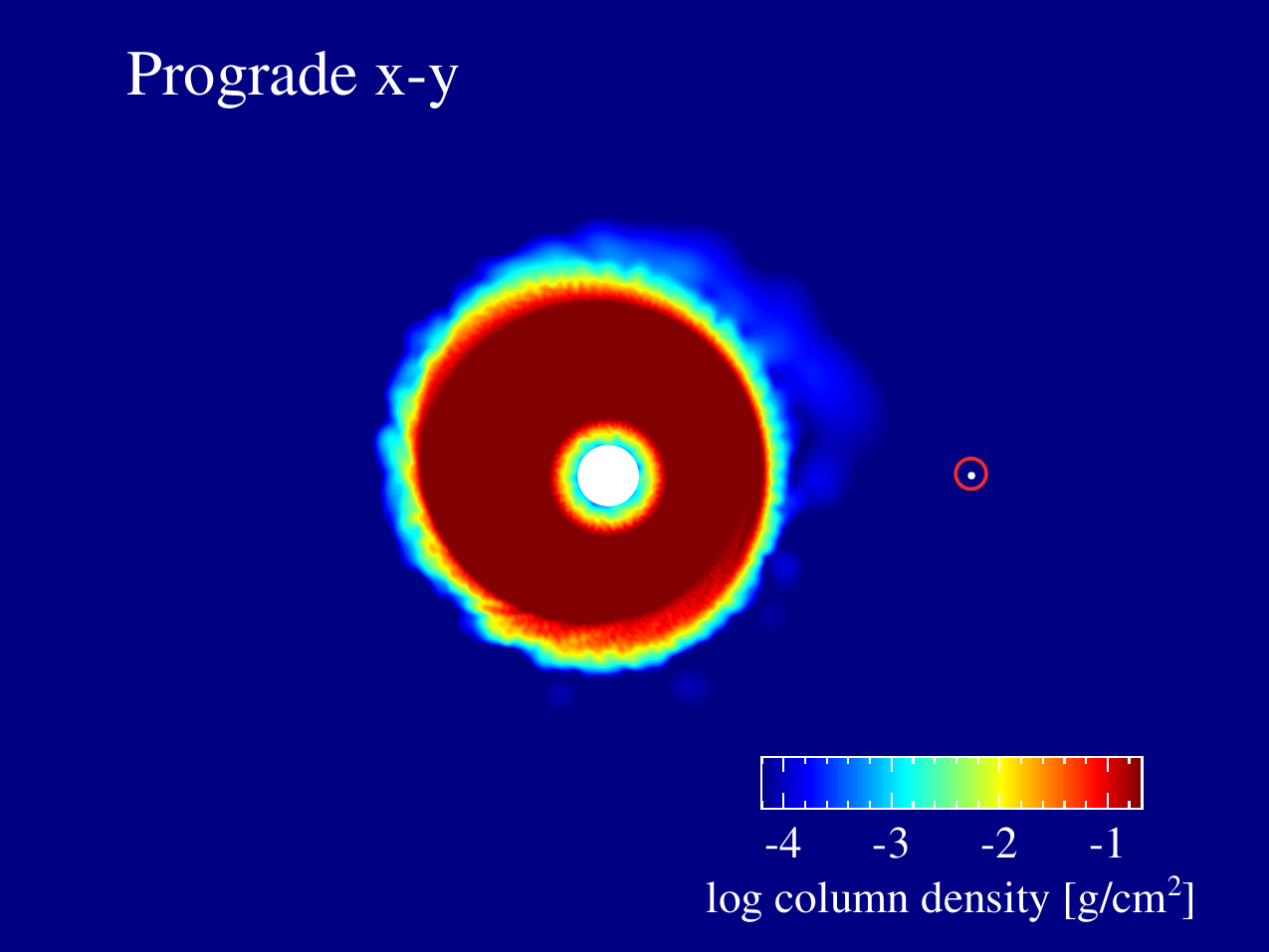}
    \includegraphics[width=0.4\textwidth]{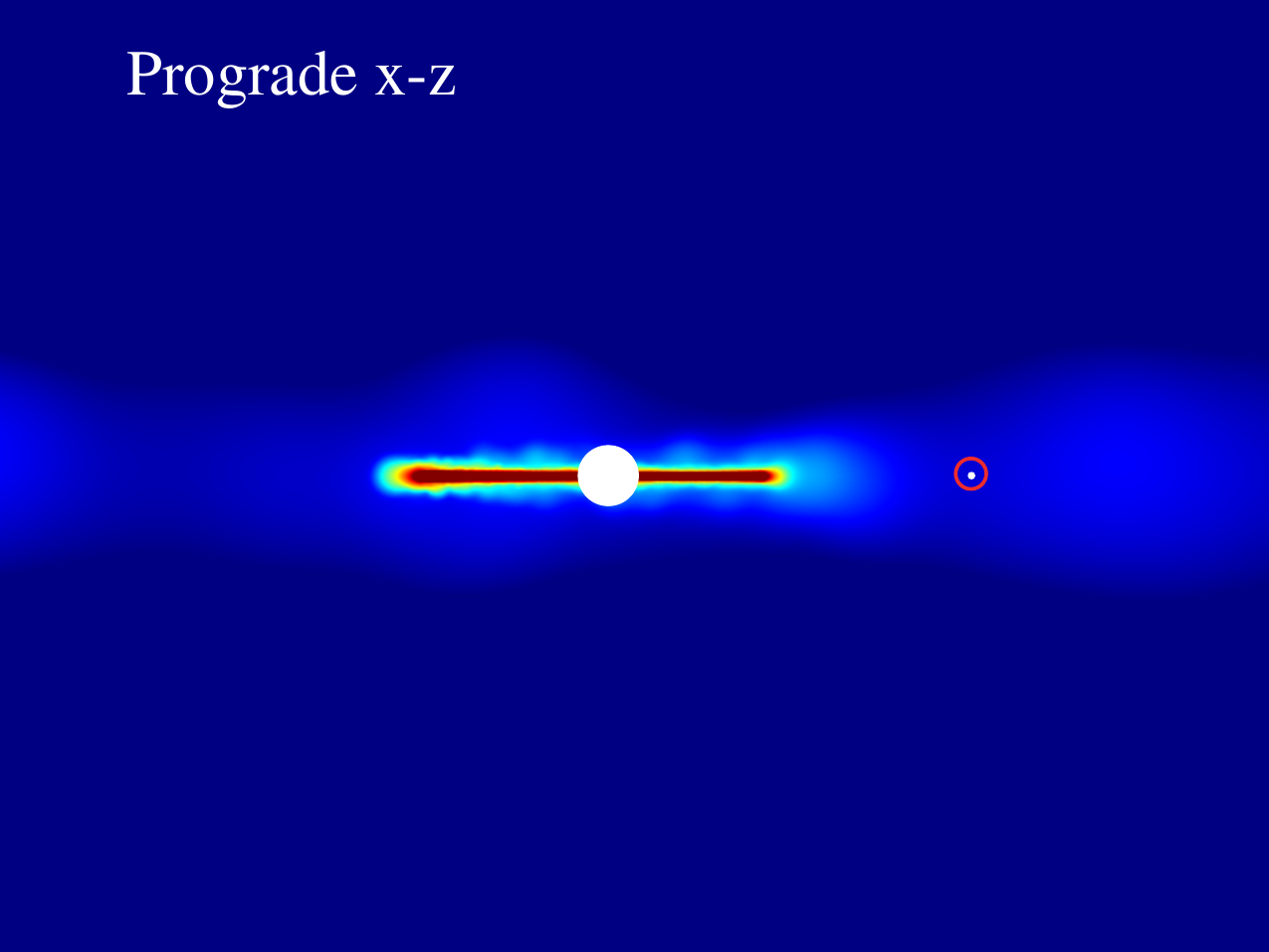} 
    \includegraphics[width=0.4\textwidth]{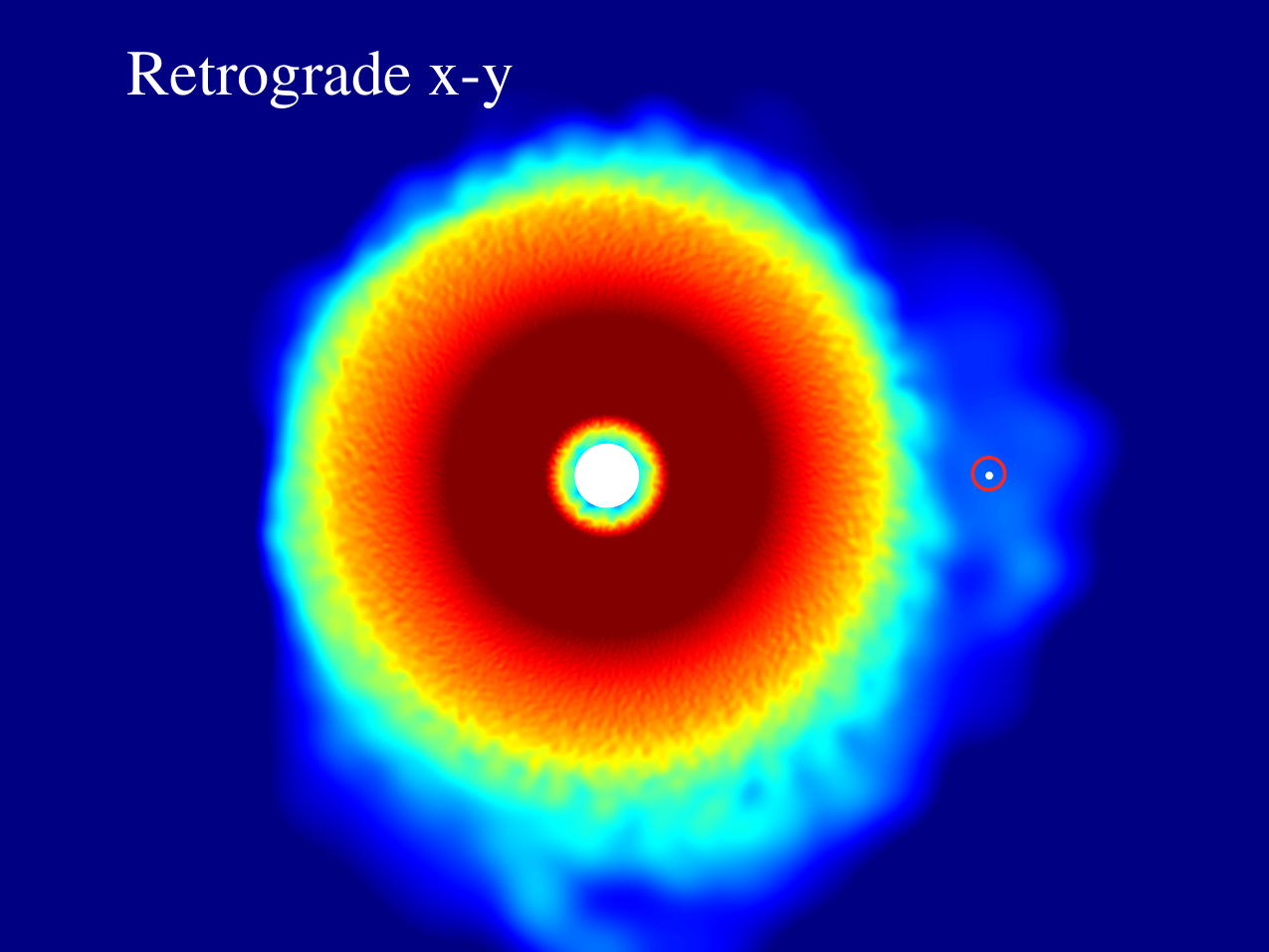}
    \includegraphics[width=0.4\textwidth]{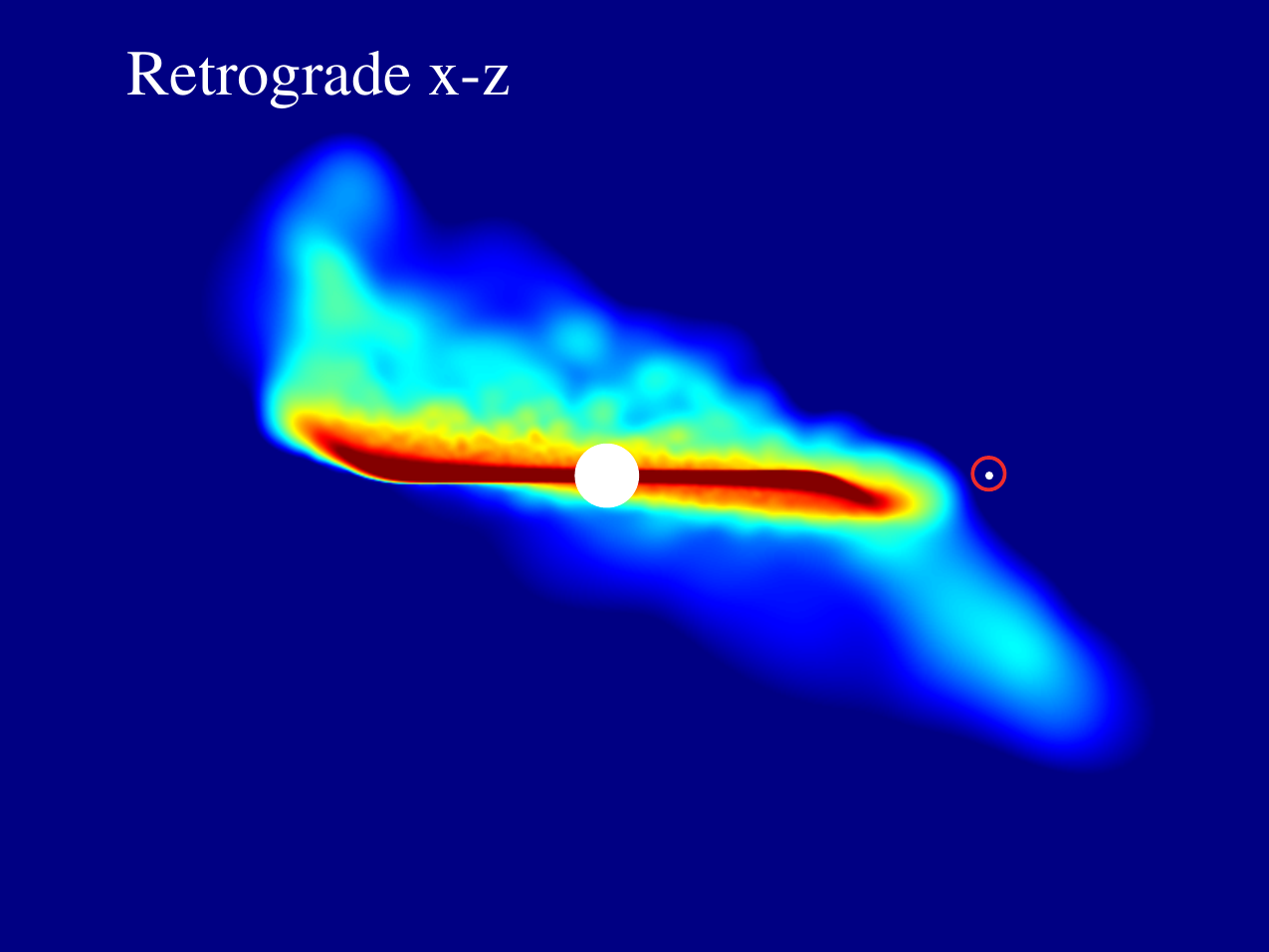}
    \caption{Column density of the Be star disc from the SPH simulations at a time of $t=41.6 \ P_{\rm orb} \,$. The top panels show the prograde disc and the bottom panels show the retrograde disc. The Be star is represented by the large white circle while the small white dot surrounded by a red circle represents the companion neutron star. The binary separation is $95 \ R_{\odot}$. The size of the white circles are scaled to the size of the sink radius. The left panels show the $x-y$ plane in which both the Be star and the neutron star orbit. The right panels show the $x-z$ plane. 
    } 
    \label{fig:splash_0s}
\end{figure*}

The orbital evolution of a prograde disc has been previously extensively studied \citep[e.g.][]{Lubow1991,Lubow1992,Murray1998,Goodchild2006}, we consider it here just for comparison to the retrograde case. We bin the particles into 100 radial bins evenly distributed between $R = 8 \ R_{\odot}$ and $R = 100 \ R_{\odot}$ and average the particle properties within each bin. The left panels of Fig.~\ref{fig:sigma_i} show the surface density and inclination to the binary orbital plane of the prograde disc as a function of radius at different times.  The prograde disc viscously expands outwards to a radius that is close to the 3:1 resonance given by 
\begin{equation}
    R_{\rm res} = 3^{-2/3}(1+ q_{\rm b})^{-1/3}a
\end{equation}
\citep{Goodchild2006}. For our parameters ($q_{\rm b}= M_{\rm NS }/ M_{*}=0.078$), the tidal truncation radius is $R_{\rm res} = 44.5 \ R_{\odot}$. The disc surface density drops off outside of this radius. There is no inclination growth relative to the binary orbital plane in the prograde disc.

Fig.~ \ref{fig:splash_0s} shows the column density of the circumprimary disc from different viewpoints at a time $t= 41.6 \ P_{\rm orb}$. The top left panel shows the prograde disc in the $x-y$ plane,  the plane of the binary orbit. The $z$-axis is along the direction of the binary angular momentum. The prograde disc becomes eccentric, reaching a maximum eccentricity of $e \approx 0.2$. The disc eccentricity growth is explained by the effects of the 3:1 resonance \citep{Lubow1991}. As the eccentricity in the disc increases, the neutron star is able to capture material during each orbital period when it passes close to the disc apastron \citep{Franchini2019b}. We see evidence of type~I outbursts for time $t\gtrsim 20 \ P_{\rm orb}$. The top right panel shows the prograde disc in the $x-z$ plane, demonstrating the coplanarity of the disc.

\subsection{Retrograde disc}
\label{retro}

The right panels of Fig.~\ref{fig:sigma_i} show the surface density, inclination relative to the binary orbital plane and nodal phase angle of the retrograde disc. Comparing the top panels of Fig.~\ref{fig:sigma_i}, we see that the retrograde disc expands to larger radii than the prograde disc. This is explained by the lack of ordinary Lindblad resonances in a circular, coplanar retrograde disc \citep{Nixonetal2011a, Lubowetal2015}. The interaction times between the orbiting companion and nearby gas in the disc are very different depending on the direction of rotation of the disc.  
The interaction time combined with the viscous expansion of the disc controls the effect of the binary on the disc. 
Therefore, there is no \textit{tidal} truncation acting on the retrograde disc due to the very short companion interaction time. Instead, the retrograde disc expands until it reaches the orbit of the neutron star. At later times in the simulation the surface density drops off with a steeper gradient around $R\approx 70 \ R_{\odot}$. This indicates there is another truncation mechanism acting on the retrograde disc related to the close interaction of the material with the neutron star. The gas particles can be either accreted on to the neutron star or ejected to circumbinary material due to this interaction \citep{Franchini2019b}.
The tidal truncation radius of the prograde disc is smaller than the truncation radius of the retrograde disc.

 Over time, the outer radii ($R \gtrsim 50 \, R_{\rm \odot}$) of the disc become significantly misaligned to the binary orbital plane. By time $t=25.05 \ P_{\rm orb}$ the inclination of the retrograde disc reaches a quasi steady state in inclination between approximately $50 \ R_{\odot}$ and $70 \ R_{\odot}$, where the inclination increases nearly linearly with radius. Beyond $70 \, R_{\odot}$ the inclination continues to increase with radius for all times. At $80 \, R_{\odot}$, the inclination reaches a maximum of approximately $i = 165^{\circ}$. The bottom right panel of Fig.~\ref{fig:sigma_i} shows the nodal phase angle (longitude of ascending node) as a function of radius for portions of the disc that are inclined less than $179^{\circ}$. Once the disc becomes tilted away from the binary orbital plane, it undergoes nodal precession driven by the binary companion \citep[e.g.][]{Larwoodetal1996, PT1995, Lubow2000,Bateetal2000}. Because the disc orbits in a retrograde direction, the nodal precession is prograde with respect to the binary orbit. The nodal phase angle increases with radius before reaching a plateau. This indicates that the disc becomes twisted due to differential precession and the outer parts of the disc precess together.

The bottom two panels of Fig.~\ref{fig:splash_0s} show the column density of the retrograde disc from a top down (left panel) and edge on (right panel) view of the binary. Unlike the prograde disc case, the retrograde disc remains circular throughout the simulation. Additionally, as visible in the edge-on view, the outer radii begin experiencing inclination growth out of the binary orbital plane and the disc becomes warped.

The disc mass decreases throughout the simulations and particles from the disc may be accreted by the neutron star or ejected into circumbinary material \citep[e.g.][]{Franchini2019}. There is less accretion onto the neutron star in the retrograde case than in the prograde case. In the retrograde case, the average accretion rate onto the neutron star over the entire duration of the simulation is approximately two orders of magnitude lower than in the prograde case. 
For the retrograde disc, more mass is ejected into the circumbinary material than accreted onto the neutron star by a factor of about $12$. In contrast, this factor is only about $0.1$ in the prograde case meaning that more material is accreted onto the neutron star than forming circumbinary material. In the prograde case there are accretion streams from the disc through which material is accreted on to the neutron star. These streams do not occur in the retrograde case.

The simulations presented here are independent of the total disc mass.  However, the quasi steady state surface density profile seen at later times is affected by the temperature profile. Therefore we have also run a simulation with an alternative value for the $q$ value given in equation~(\ref{c_s}).  Increasing the value of $q$ creates a steeper drop off in the steady-state surface density. With $q = 1$ we find a similar magnitude of tilting to the main results of this work (where $q = 0.5$).


\section{Retrograde disc tilt instability}
\label{sec:test_part}

\begin{figure}
	\includegraphics[width=\columnwidth]{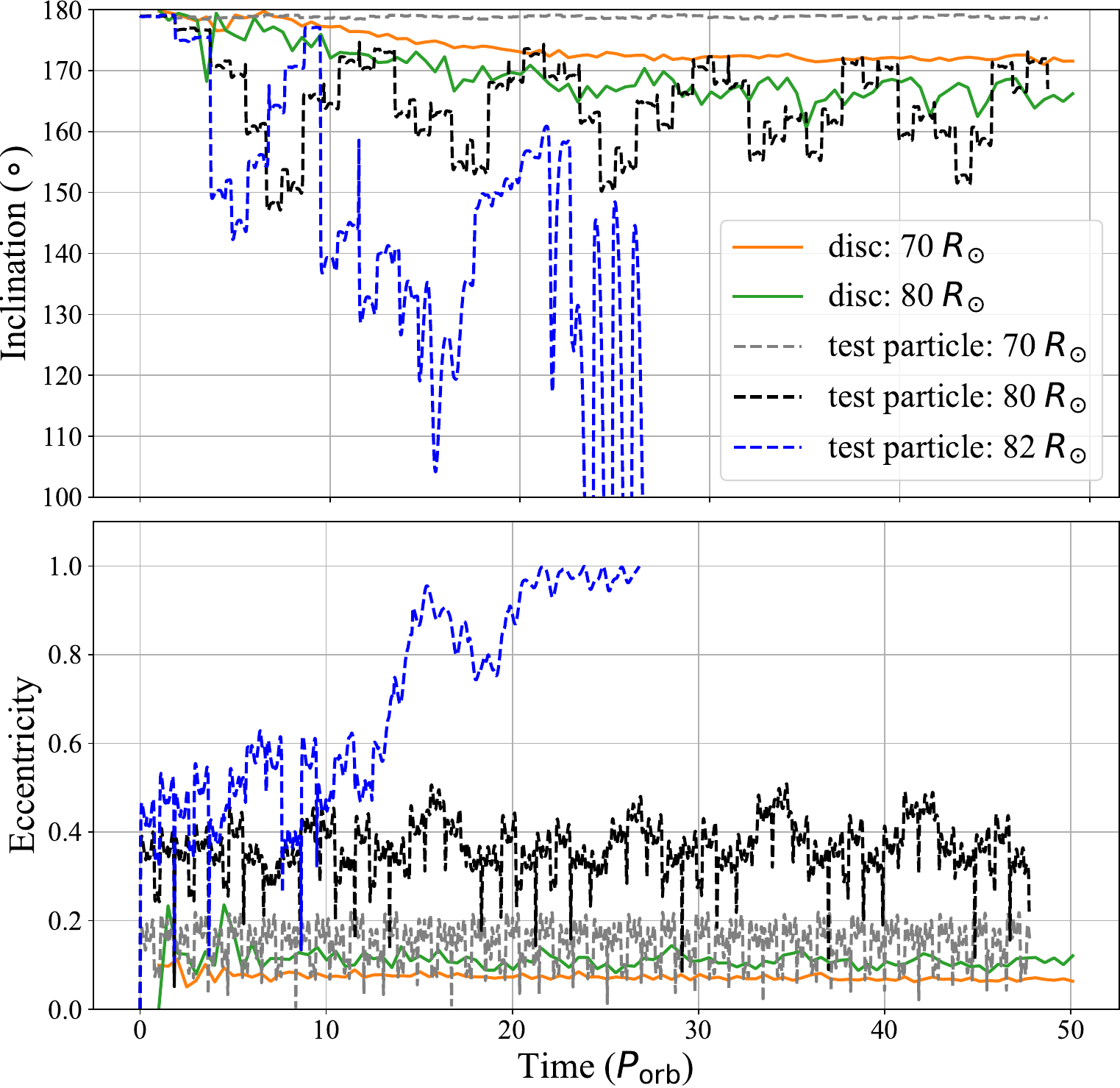}
   \caption{Test particle inclination relative to the binary orbital plane (upper panel) and eccentricity (lower panel) as a function of time. The  initial orbital  radius is $R=70$ (gray), $80$ (black) and $82 \ R_{\odot}$ (blue). The retrograde disc properties (from Section~\ref{retro}) are shown at $R=70 \ R_{\odot}$ (orange) and $80 \ R_{\odot}$ (green). The blue line is truncated at the time when the particle is ejected (eccentricity $>1$).  }
   \label{fig:e_t_rebound}
\end{figure}

In order to better understand the cause of the warping in the disc simulation, we use the $n$-body integrator, {\sc Rebound} \citep{rebound,reboundias15} to examine retrograde test particle orbits around the Be star. We simultaneously simulate test particles at varying initial separation. These are massless test particles which do not interact with each other. The particles are placed in a retrograde Keplarian orbit which is coplanar with respect to the binary orbit.
Fig.~\ref{fig:e_t_rebound} shows the inclination relative to the binary orbital plane and eccentricity of the test particles at an initial separation of $R=70$, $80$, and $82 \ R_{\odot}$ respectively. The figure shows that particles interior to $82 \, R_{\odot}$ experience large but chaotic oscillations in inclination and eccentricity. At $R \gtrsim 82 \, R_{\odot}$, the inclination instability becomes strong enough to eject the test particles. This indicates that the inclination instability is stronger at radii close to the neutron star orbit. 
The retrograde disc inclination at $R=70$ and $80\,\rm R_\odot$ is also shown in Fig.~\ref{fig:e_t_rebound} for comparison. In the case of a disc, material can exist in the exterior region to $82 \ R_{\odot}$ (where test particles are ejected) because there is radial communication through the disc that stabilises the outer parts. The radial communication is likely dominated by the effects of viscosity rather than pressure because we choose a large value ($\alpha \approx 0.3$) for the \cite{SS1973} viscosity parameter \citep[e.g.][]{Nixon2016}. 

The vertical instability of particle orbits close to a perturber is indicative of a possible inclination instability of a fluid disc. However, the particle instability is dominated by the effects of incoherent motions that differ between nearby particles. To model a disc tilt, it is important include collective effects such as viscosity and pressure that cause the disc to respond coherently and limit the tilt near the resonance \citep[e.g.][]{Meyer1987}.

A disc around one component of a circular orbit binary that undergoes nodal precession in the presence of dissipation moves towards retrograde alignment \citep[e.g.][]{Papaloizou1995, Lubow2000, Bate2003}. Therefore there is a balance between the viscosity that acts to align the disc to retrograde coplanar, and the tilting mechanism. A larger tilt would therefore be expected with a smaller viscosity.

Ordinary Lindblad resonances due to the companion are not present in a retrograde disc. At such resonances we have
\begin{equation}
    \Omega = \frac{m \Omega_{\rm b}}{m\pm 1},
\end{equation}
for azimuthal wavenumber $m$ which is a positive integer and $\Omega = \sqrt{G M_* / R^3 }$ is always positive, which corresponds to prograde orbits. The angular velocity of the binary companion is $\Omega_{\rm b}=2\pi / P_{\rm orb}$. Lindblad resonances arise for tilted discs due to their interaction with the tidal potential of the binary. These resonances satisfy the condition that
\begin{equation}
    \Omega = \frac{m \Omega_{\rm b}}{m\pm 2}.
\end{equation}
\citep{Lubow1992b}. For $m=1$, a retrograde resonance is possible for $\Omega=-\Omega_{\rm b}$. At such a resonance a fluid disc can be tilt unstable. 
Although we simulate an initially coplanar disc, small numerical departures from exact coplanarity of the disc midplane inevitably occur that can be amplified to large values by an exponentially growing instability.

Unfortunately there is a singularity in the companion potential in a model involving an extreme binary mass ratio that occurs at a radius $\Omega=-\Omega_{\rm b}$ and so the stability analysis in \cite{Lubow1992b} cannot be directly applied. The disc does not extend to the resonance, but the resonance has some width over which it could act. Near resonant effects may then play a role. This provides a possible explanation for the tilting of a retrograde disc as well as the inclination of the retrograde test particle orbits. This explanation should be explored further. An estimate for the width of the resonance can be found from the outer radius of the disc at the time it first experiences inclination growth. In the top right panel of Fig. \ref{fig:sigma_i} the line corresponding to $5.01 \ P_{\rm orb}$ has a small rise in the inclination at the very outer edge (at approximately $65 \ R_{\odot}$). At this point, there is no influence from exterior material which may cause the inclination growth, indicating this radius is where the resonance becomes strong enough to influence the disc.

\section{Conclusions}
\label{concs}

We have found that a retrograde disc orbiting around one component of a binary is unstable to tilting. With hydrodynamical simulations we have modelled coplanar and circular binaries. For a retrograde disc we find that the disc becomes tilted away from retrograde coplanar with the largest inclination growth relative to the binary orbital plane occurring in the outermost radii. We have examined test particle simulations and seen significant inclination growth in their orbits that increase in magnitude with radius from the primary star. The inclination growth may be due to a retrograde Lindblad resonance occurring at $\Omega=-\Omega_{\rm b}$ which acts over some width. In a disc, the large tilt in the outermost radii is spread over the radial width of the resonance and propagates through the disc resulting in inclination growth away from the location of the resonance.

This tilt instability has implications for accretion in Be/X-ray binaries. The strong warping in the outer parts of the Be star disc means that there is less material in the binary orbital plane close to the companion. There is therefore less accretion on to the neutron star in a retrograde system than would otherwise be expected. Since X-ray outbursts are driven by accretion on to the neutron star, these outbursts are expected to be weaker in a retrograde system.

\section*{Acknowledgements}

We thank an anonymous referee for a thorough review and providing useful comments. We thank Daniel Price for providing the phantom code for SPH simulations and acknowledge the use of SPLASH \citep{Price2007} for the rendering of the figures. The test particle simulations made use of the REBOUND code which is free to download at http://github.com/hannorein/rebound. Computer support was provided by UNLV’s National Supercomputing Center. We acknowledge support from NASA through grant 80NSSC21K0395.

\section*{Data Availability}

The data underlying this letter will be shared on reasonable request to the corresponding author.


\bibliographystyle{mnras}
\bibliography{long_cite}






\bsp	
\label{lastpage}
\end{document}